\documentclass[a4paper]{jpconf}
\usepackage{graphicx}
\usepackage{epsfig}
\usepackage{amsmath}
\usepackage{fancyhdr}
\renewcommand{\thefootnote}{\fnsymbol{footnote}}

\begin{document}
\title{Critical properties of the antiferromagnetic Ising model on rewired square lattices}
\author{Tasrief Surungan$^1$, Bansawang BJ$^1$ and Muhammad Yusuf$^2$}
\address{$^1$Department of Physics, Hasanuddin University, Makassar, South Sulawesi,
 90245, Indonesia
$^2$Department of Physics, Gorontalo State University, Gorontalo, Gorontalo 96128, Indonesia}
\ead{tasrief@unhas.ac.id}
\begin{abstract}
The effect of  randomness on  critical behavior  is a crucial subject 
in condensed matter physics due to the the presence of impurity in any real material. 
We presently probe the critical behaviour of the  antiferromagnetic (AF) Ising model on 
rewired square lattices with random connectivity.
An extra link is randomly added to each site of the square lattice
to connect the site to one of its  next-nearest neighbours,
thus having  different number of connections (links).   
Average number of links (ANOL) $\kappa$ is fractional,  varied from 2 to 3, where
$\kappa = 2$ associated with the native square lattice. The rewired lattices
possess abundance of triangular units in which spins are frustrated
due to AF interaction.  The system is studied by using Monte Carlo method with
Replica Exchange algorithm. Some physical quantities of interests
were calculated, such as the specific heat,
the staggered magnetization and the spin glass order parameter
(Edward-Anderson parameter).  We investigate the role played by the randomness
in affecting the existing phase transition and its interplay 
with frustration to possibly bring any  spin glass (SG) properties.
We observed  the low temperature magnetic ordered phase (N\'eel phase) preserved 
up to  certain value of $\kappa$ and no indication of SG phase for any value of $\kappa$.
\end{abstract}
\section{Introduction}
The cooperative phenomena are ubiquitous in nature, driven by the presence
of coupling interaction between each individual constituent of  materials
\cite{Ising,Sachdev}.  This  is exemplified by the spontaneous magnetization 
by which  certain magnetic material is  entering  a ferromagnetic phase below 
its transition temperature $T_c$ (Curie temperature) and  able to attract the nearby 
metals surrounding it. The high temperature  disordered phase and the
cooperative ordered phase at low temperature  are separated by $T_c$.
A phase transition   is essentially rendered by 
the competition between the external fields, such as temperature, 
which tend to destroy the order and the coupling interaction which  tends to create 
order.

Based on the nature of free energy function, a  phase transition is 
grouped into the first and the second order phase transition\cite{Ehren}.
The transition is called as first order
if the first order derivative of the free energy of the system
is discontinuous.  If the derivative is continuous,
the transition is second order, also called continuous phase
 transition, at which no latent heat involved in the course of transition 
and no co-existing phases\cite{Sachdev, Nishimori}. A firm example of these is the 
phase transition experienced by  PVT systems  above  the critical point\cite{Reichl}.
The  phenomena nearby critical point are known as critical phenomena, and
 affected by limited number of controlling  variables such as the type of coupling 
interaction, symmetry of the microscopic constituents (e.g., spins)
and spatial dimensionality.  Critical phenomena are characterized by 
a set of critical exponents which are in general universal, where different 
systems may be grouped in the same universality class, i.e.,  having similar 
value of exponents.

The type of phase transitions and  universality classes are subject to the controlling 
variables and affected by the presence of randomness.  The presence of randomness is a 
crucial aspect in the study of phase transition; and has long been intensively 
investigated.  A randomness can be realized in a variety of forms such as defect 
(site vacancies) and diluted bonds.  
The effect  of bond dilution for system  undergoing  Kosterlitz-Thouless (KT) transition 
has been investigated and found that the KT phase preserved so long as the bonds
of the lattice are percolated\cite{Surungan2005}.
According to Harris criterion\cite{Harris},
randomness is relevant (able to change the type or destroy the order or
alter the universality classes)  for the pure system experiencing second order 
phase transition with specific heat exponent $\alpha > 0$,
and irrelevant for $\alpha < 0 $; otherwise it is marginal.

In this study we investigated the role played by the
connectivity randomness in affecting critical properties
of the Ising model on rewired square lattice with AF couplings. The pure 
system of this model is well known to experience second order phase transition 
with $\alpha > 0$\cite{Yang}. The crucial point is that the AF couplings bring
 geometrical frustration due to the presence of diagonal neighbours
randomly added to the  square lattice. 
A spin is  frustrated  if it can not find a unique orientation 
in interacting with its neighbors to stay in a minimum energy state\cite{Diep}.
There will be an interplay between randomness and the geometrical 
frustration, which may affect the existing low temperature 
ordered phase (the N\'eel phase) and could  possibly lead to  SG ordering\cite{Cannella}.
In fact, various spin models on rewired lattices with AF interaction 
have  been probed  in connection to SG problem\cite{CCP2014,ICTAP2014,CCP2015}. 
Previously it was found that no finite temperature Ising SG phase 
observed on rewired square lattice  for $\kappa =3$\cite{Palembang}.
The study did not pay attention to the role played by the randomness
in affecting the existing order phase.
To systematically study this topic,  here we consider fractional values
of $\kappa$, ranging from 2 to 3. We probed  
the existing second order phase transition and 
the N\'eel phase at low temperature  due to the presence of randomness and frustration.
In particular we searched for the disappearance of AF orderings due to the 
increasing value of $\kappa$.
\begin{figure}
\begin{center}
\includegraphics[width=0.50\linewidth]{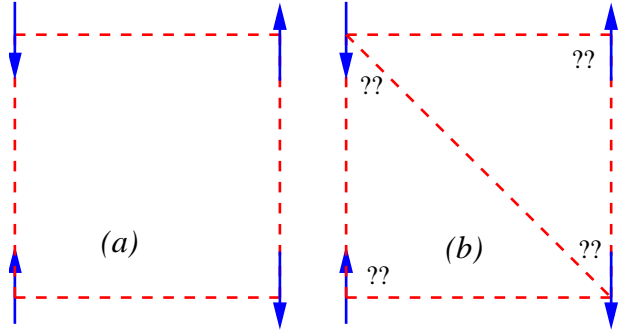}
\vspace{-0.2cm}
\caption{(a) Unfrustrated spins and (b) geometrical frustrated spins
on rewired plaquettee.  The AF couplings
are represented by the dashed lines. }\label{Fig01}
\end{center}
\vspace{-0.9cm}
\end{figure}

The subsequent parts of the paper is organized
as follows: Section II explained the model and
method. Results and discussion are presented in Section III. Section
IV is devoted to the summary and concluding remarks.

\section{Model and Simulation}
The model is written in the following Hamiltonian
\begin{equation}
H =  - \sum_{\langle i j \rangle} J_{ij} s_i s_j
\end{equation}
where $J_{ij}$ is the coupling interaction between the Ising
spin $s$ occupying respectively the site $i$-th and
$j$-th of the lattice.  As we probed a purely AF system,
 $J_{ij} = J < 0$. The summation is carried out over all 
pairs of spins directly connecting.
For  regular lattices such as the square and the cubic lattice, 
all spins possess similar number of neighbours,
associated with a fixed coordination number. Currently, 
we consider rewired square lattices 
with fractional connectivity, i.e., the average number of links (ANOL), $\kappa$,
varies from  2 to 3.  
A rewired lattice is considered as a quasi-regular structure as the translational
symmetry partially preserved.  Unlike a random structure
or a scale free network\cite{Barabasi}, where the notion of spatial
dimension is obscure, for a quasi-regular lattice dimensionality remains. 
Our procedure of constructing the lattice is by  randomly connecting
each site to one of the diagonal neighbors of the site.
There are 4 native nearest neighbors of the original square lattice;
after being  rewired, where one link is randomly added to each site, 
 we obtain a lattice with fractional ANOL. 
Each individual site may have in minimum of 4 and maximum of 7 neighbors.  

The randomly rewired lattices with AF couplings inherits the main
ingredients of SG system and is considered as a new type of 
SG model, called non-canonical SG \cite{CCP2014,CCP2015,Bartolozzi,Herrero}.
It is  distinguished  from the canonical model as the type of randomness 
and frustration are not due to the presence of both FM and AF couplings. 
In a  canonical model, both FM and AF coupling exist, while
for non-canonical model only AF couplings exist,
symbolically indicated by the dashed lines in Figure \ref{Fig01}.
The larger the value of $\kappa$ the larger the number of triangular units.
As shown in Fig. \ref{Fig01}, spins in AF triangular unit are frustrated. 
If we further  add some extra links the degree of frustration
of the system increases. The number of triangular units
determines the degree of frustration.
This is analogous to the canonical SG system with various distribution
 FM and AF couplings, resulting SG phase diagram with Nishimori line\cite{Nishi83}.

As a standard method in statistical mechanics, we used  Monte Carlo (MC) 
simulation to compute the physical quantities.
Due to the complexity of the energy landscape, which may
be difficult to tackle using the the conventional  Metropolis
algorithm, here we used the Replica Exchange (RE) algorithm\cite{Hukushima}.
It is a powerful MC algorithm  introduced for  studying complex problems
 such as randomly frustrated system, in particular to overcome the problem of slow dynamics.
This problem is commonly found in complex system
because of the presence of local minima in the energy landscape,
where a random walker can easily  get trapped at
certain local minimum. The basic idea of this algorithm
is to extend the conventional Metropolis
algorithm by duplicating  the original systems into several replicas.
Each replica is then  independently simulated using a standard
Metropolis algorithm. As each replica belongs to a heat bath with
 certain temperature, then the whole result of the simulation is obtained
by combining each individual result.
During the computation, two replicas associated 
with to neighboring temperatures  are exchanged.
This is the essential trick for the RE algorithm,
by which a random walker is able to escape from any  local minimum.

If we begin the simulation  with a number $K$ of replicas; where
every replica is in equilibrium with a heat bath of a corresponding 
temperature $T=1/\beta$, then the probability distribution of observing
the whole system in a state
 $\{ X \} =
\{ X_1,X_2, \dots, X_M\}$
is given by,
\begin{equation}
 P(\{ X, \beta\}) = \prod_{m=1}^{M} \tilde{P}(X_{m},\beta_{m}),
\end{equation}
with
\begin{equation}
\tilde{P}( X_m, \beta_m) = Z(\beta_{m})^{-1} \exp(-\beta_{m} H(X_{m})),
\label{equil}
\end{equation}
Here, the partition function of the m-$th$ replica is represented by $Z(\beta_m)$. 
Next, we can  define an exchange matrix between two replicas,
 $W(X_m,\beta_m| X_n,\beta_n)$. This matrix is  the probability
to swap the configuration $X_m$ of  $\beta_m$
with the configuration $X_n$ of $\beta_n$.
In order to maintain  the entire system at equilibrium,
the detailed balance condition is imposed on the transition matrix

\begin{eqnarray}
P(\{ X_m, \beta_m \},\ldots, \{ X_n, \beta_n \},\ldots )\cdot
 W(X_m,\beta_m| X_n,\beta_n) \nonumber \\ = P(\{ X_n, \beta_m \},
\ldots, \{ X_m, \beta_n \},\ldots )
\cdot W( X_n,\beta_m | X_m,\beta_n), 
\end{eqnarray}
along with Eq.~(\ref{equil}), so that we obtain
\begin{equation}
\frac{ W( X_m,\beta_m | X_n,\beta_n)}{ W( X_n,\beta_m | X_m,\beta_n)}=\exp(-\Delta),
\end{equation}
where $\Delta=(\beta_{n}-\beta_{m})(H(X_{m})-H(X_{n}))$.
Based on  this constraint, we can choose the coefficient of matrix 
according to the standard Metropolis algorithm which gives
the following
\begin{equation}
W(X_m,\beta_m| X_{n},\beta_n)=\left \{  \begin{array}{ccc} 
1 & {\rm if} & \Delta<0, \\ 
\exp(-\Delta) & {\rm if} & \Delta>0.
\end{array} \right.
\label{trans}
\end{equation}
Since the ratio of acceptance ratio exponentially decays 
with $(\beta_n-\beta_m)$,  the swap is carried out only between
teo adjacent temperatures, i.e., the terms $W(X_m,\beta_m| X_{m+1},\beta_{m+1})$.
\begin{figure}
\includegraphics[width=0.51\linewidth]{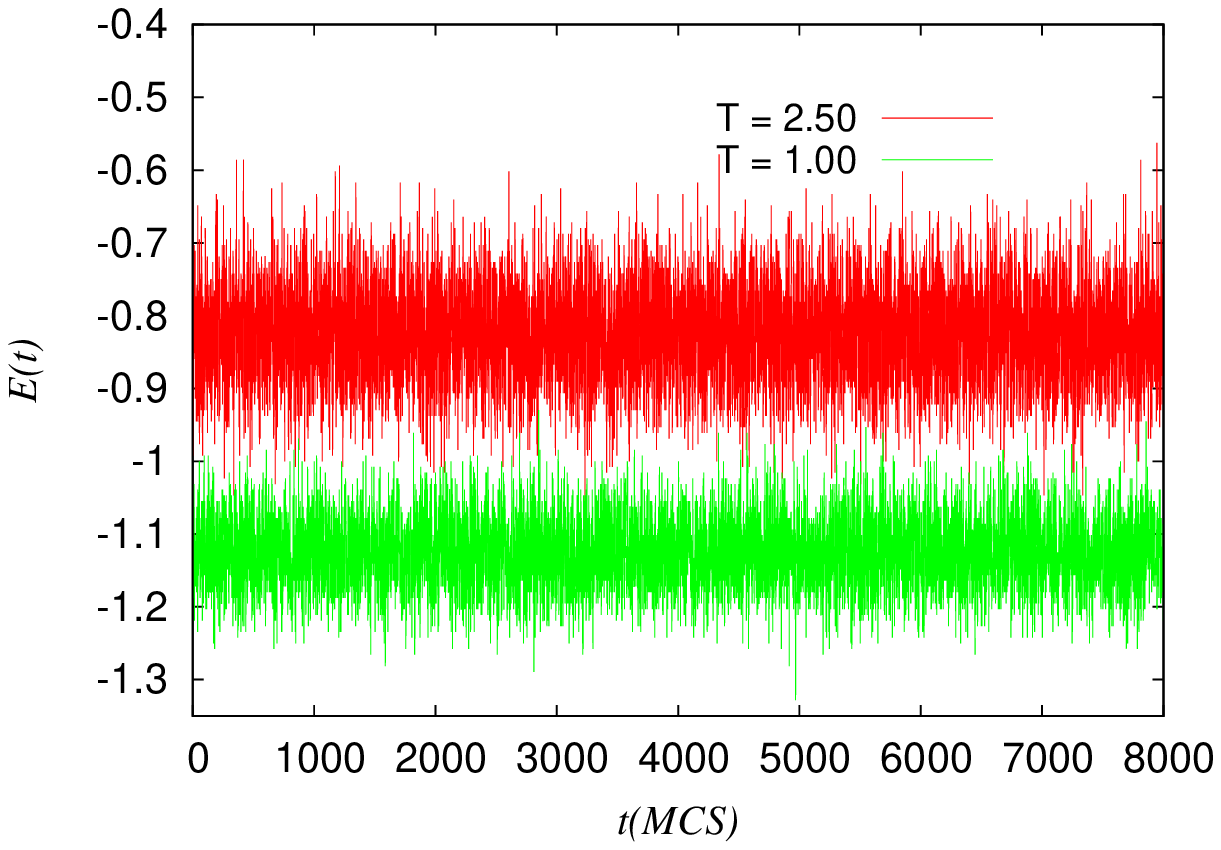}
\includegraphics[width=0.51\linewidth]{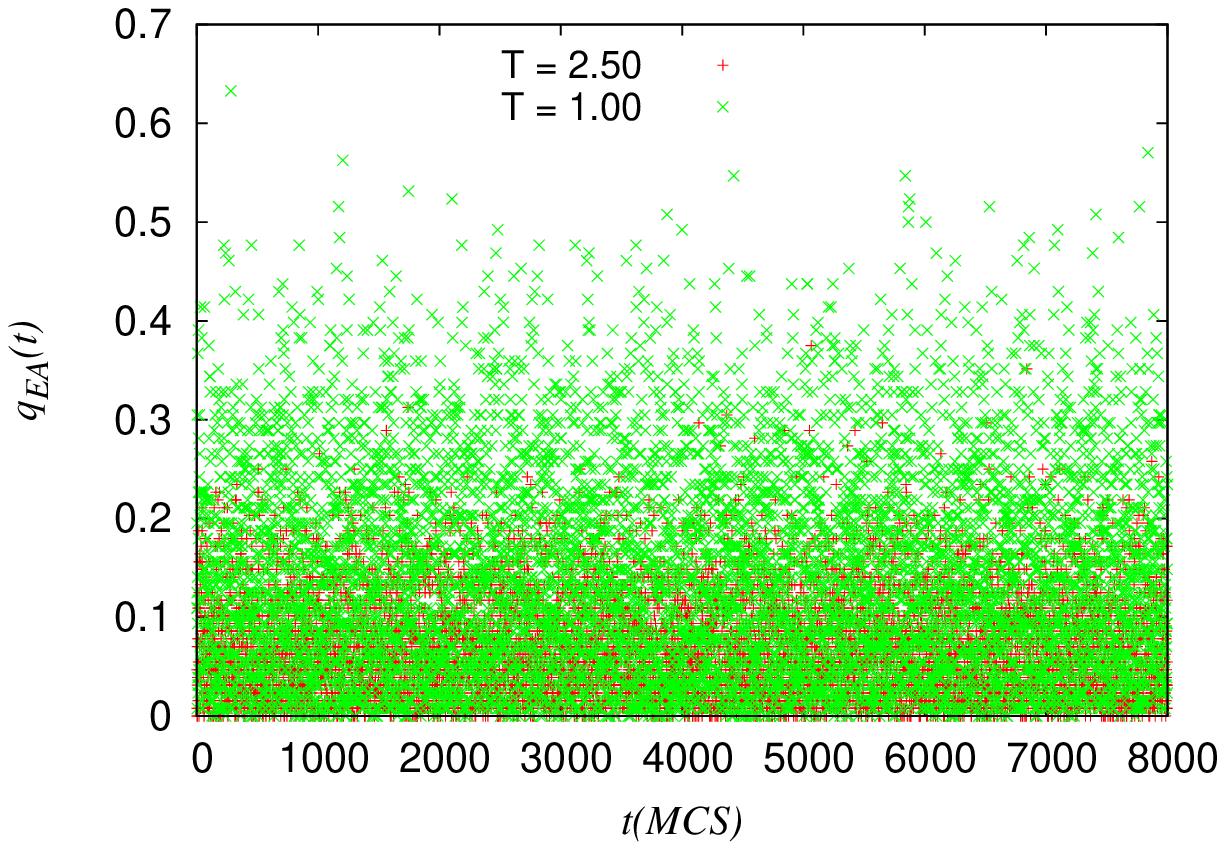}
\caption{(Color online) Time series of enegy (left); and Edward-Anderson order 
parameter (right)  at $T=2.50$ and $1.00$ for system size $L=16$ with  $\kappa$=3.0.}
\label{TS24}
\vspace{-0.3cm}
\end{figure}

In carrying out the simulation, we define a step of  MC (MCS)
as updating  each spin once for each replica, either randomly
or consecutively.  After enough MCSs,
we swapped each pairs of configurations belonging to 
neighboring temperatures according to 
the probability according to Eq. \ref{trans}.
For example, we  swap replicas $X_1$ and and $X_2$, 
based on the probability defined Eq. \ref{trans}. 
Then, two consecutive swaps are assigned
as even and odd and. For the even swap,
we exchange the pairs of replicas $X_2$ and $X_3$;
$X_4$ and $X_5$; $\cdots$; $X_{M-1}$ and $X_M$;
while for an odd one we exchange $X_1$ and
$X_2; \cdots X_{M-1}$ and $X_{M-3}$.

For a certain realization of connectivity, we started from
a random spin configuration. Then, we  equilibrated the system with
$10^4$ MCSs before doing calculation with total of $5 \times 10^4$
samples for each temperature. This data make up time series
of corresponding physical quantities, from which we extract the thermal averages.
Several additional MCSs between two samples  were taken in order
to avoid high time-correlation.  We simulated the system with reasonable number 
of temperatures, so that covering  the high and low temperatures.
The temperature dependence of a corresponding
physical quantities, apart from being averaged over samples, it 
is also averaged  over many lattice connectivities for each
system size. This is a standard procedure  in  MC simulation in dealing 
with  random systems. In the next section, we present and discuss our results.

\section{Results and  Discussion}
\subsection{Time series of energy and specific heat}
We have studied AF Ising model  on rewired square lattices
with various values of $\kappa$. 
Several linear sizes, $L=16,24, 32$ and $48$,
were simulated, with $N=L^2$ is the total number of sites (spins).
Since  the periodic boundary condition is imposed,
each site of  the native square lattice has the same number of nearest neighbors.
For each system size we took many realizations
of connectivities, then average the results over the number of realizations.
Each realization corresponds to a particular connectivity
distribution which is randomly generated.
As a random system, results of a particular realization
tend to be  different from that of the other.
Therefore, for a reliable result (better statistics), we took reasonable number 
of realizations.  In previous studies on various randomly frustrated 
systems\cite{Bartolozzi,CCP2014,ICTAP2014,CCP2015},  1000 realizations
were taken, here  we took 500 realizations.

In MC simulation, time corresponds to a series of MCSs.  We perform $M$ MCSs
for each temperature and take a number ($N_s$) samples  out of $M_s$.
To check whether the system is well equilibrated,   we evaluated  the
energy time series of the system, from which
the average energy and specific heat can be extracted.
The presence of any existing phase transition could 
be signified in the temperature dependence of specific heat.
To make sure the system is well equilibrated we perform
enough initial MCSs, in the order of 10$^4$ MCSs, before doing measurement.

\begin{figure}
\includegraphics[width=0.51\linewidth]{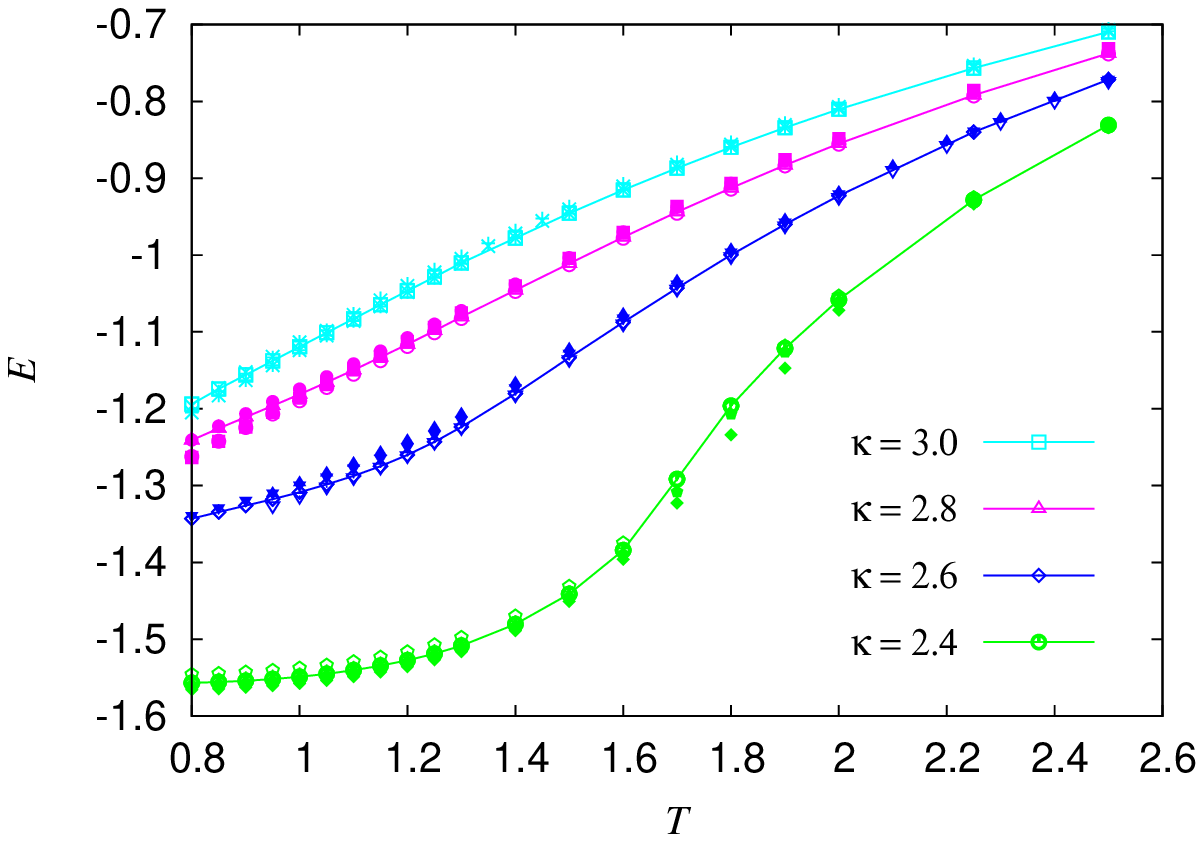}
\includegraphics[width=0.51\linewidth]{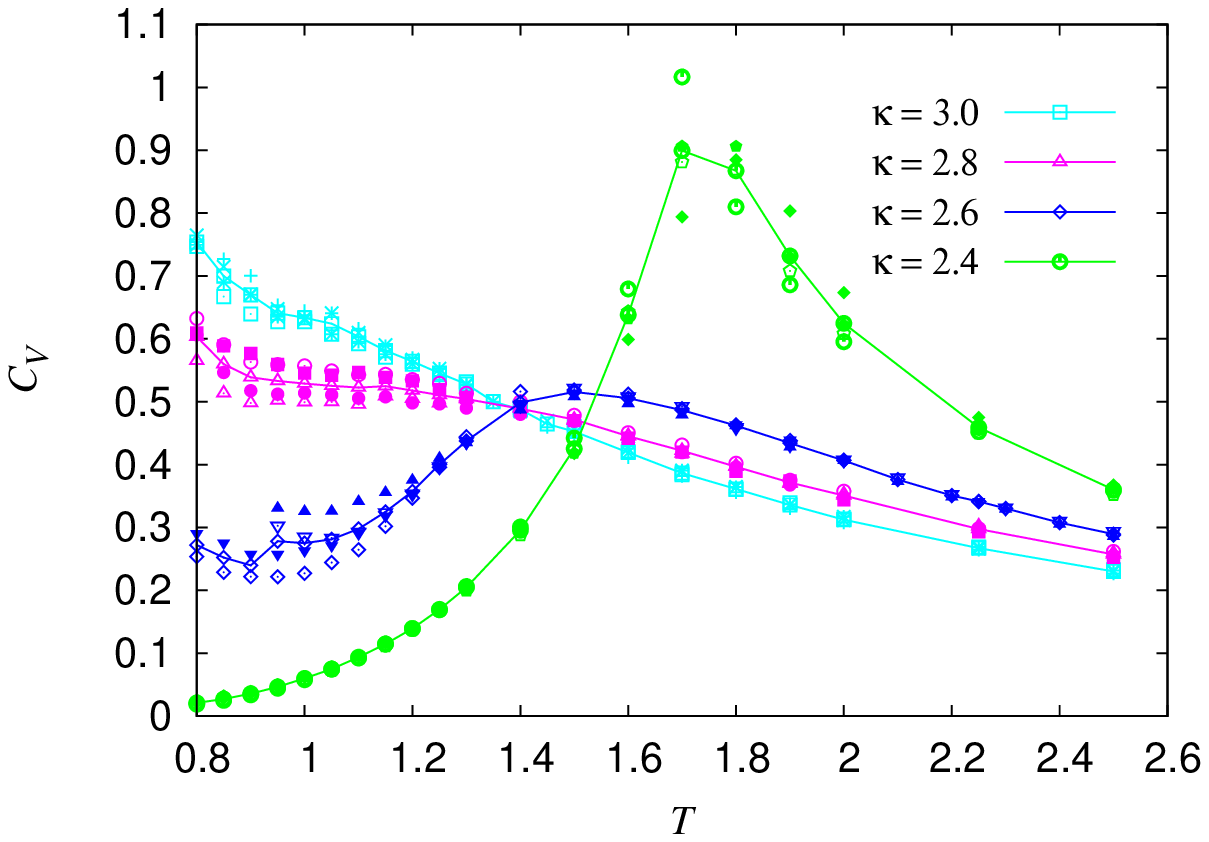}
\vspace{-0.4cm}
\caption{Temperature dependence of energy (left) and specific heat
(right) of lattices with different values of $\kappa$. As shown, for
 $\kappa=2.4$, the specific heat is diverging, signifying
a presence of second order phase transition.}
\label{spht}
\vspace{-0.1cm}
\end{figure}

The energy time series (ETS) of two different temperatures
were taken, i.e., T = 2.50 and T = 1.00, for linear size $L = 16$
of rewired square lattice with $\kappa = 3.0$.
As shown in Fig. \ref{TS24}, both plots indicate that system has achieved
equilibrium state after a number of preliminary MCSs.
Fluctuation at smaller system is more pronounced  compared
to larger system size.  We extracted two quantities from
ETS, namely the ensemble average of energy,
$\langle E \rangle =  \frac{1}{N} \sum_N E_i$, and
the specific heat  defined as follows
\begin{equation}
C_v = \frac{N}{kT^2} \left( \langle E^2  \rangle  - \langle E \rangle^2\right)
\end{equation}
where $N$ and $k$ are respectively the number of spins
and Boltzmann constant. These quantities are shown
in Fig. \ref{spht}(a) and  Fig. \ref{spht}(b). Each curve
in the  plot of average energy $\langle E\rangle $ corresponds to
a system with certain  $\kappa$.

The specific heat plot shown in Fig. \ref{spht} possesses a clear diverging peak
 for lattice with $\kappa$=2.4.   This is a strong indication of   the existence of
phase transition. Whether this transition is related to SG phase or not can be resolved 
from the data of order parameter, the EA parameter and the staggered magnetization 
$M_s$  which is presented in the next sub-section. The reason of calculating $M_s$
is due to the fact that that the original square lattice with AF coupling, before
being rewired, does experience conventional  magnetic phase transition with characterized 
by the finite value of staggered magnetization at low temperature. 
\subsection{Order Parameters}
To probe the interplay between randomness and geometrical frustration
in affecting the existing AF magnetic order phase, we calculated two corresponding
order parameters, the staggered magnetization  and the overlapping parameter, also known as 
Edward-Anderson (EA) order parameter\cite{EA}. The former is the quantity
to characterize the low temperature AF order phase, known as N\'eel phase;
while the latter is to search for the existence SG phase.  At very low temperature, 
the AF system on pure square lattice will have perfect N\'eel phase
which is characterized as follows
\begin{equation}
M_s = \sum_{i} (-1)^i M_i; ~~~~ M_i = \sum_{j\epsilon \{i\}} s_j
\end{equation}
where $M_i$ is the magnetization of the sub-lattice$i$-th,
$j$ is the spin index, depending on the sub-lattice. As  here are two sub-lattices; 
$j$ is running from 1 to N/2; where N is the number of sites.

The EA  parameter is defined as the following
\begin{equation}
q=q_{EA} \equiv \left[ \langle |\frac{1}{N}\sum_{i} s_i^{(\alpha)} s_i^{(\beta)}| 
\rangle \right]_{\rm av},
\end{equation}
where $s_i^{\alpha}$ and $s_i^{\beta}$ are two spin configurations with  
similar connectivity structure.  This parameter is more like a trick in order 
to capture the condition of frozen configuration. Certainly, it
 is  only relevant for numerical simulation and not relevant for experiment.
It is not found in real material
as  almost impossible to obtain two duplicated  systems with exactly the same 
realization of randomness.  
In experiment, the frozen state is measured directly by observing
how the configuration change with time, in other words, one measures the
time correlation of the spin configuration. 
The EA parameter for the Ising case is much simpler 
compared to the Heisenberg model which involves tensor product \cite{CCP2014,ICTAP2014}.
This  parameter is finite if the system is in the phase of
 frozen random spin configuration and vanishes otherwise.

\begin{figure}[h]
\includegraphics[width=0.51\linewidth]{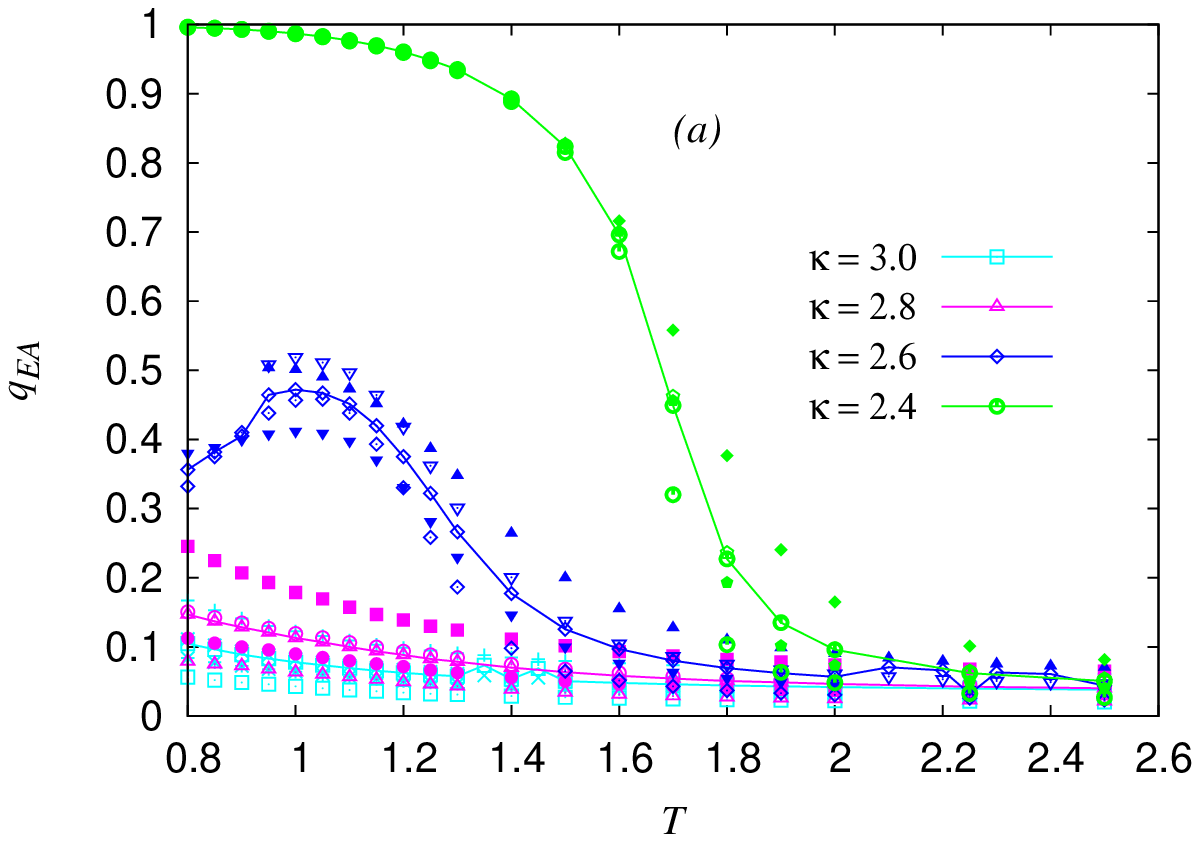}
\includegraphics[width=0.51\linewidth]{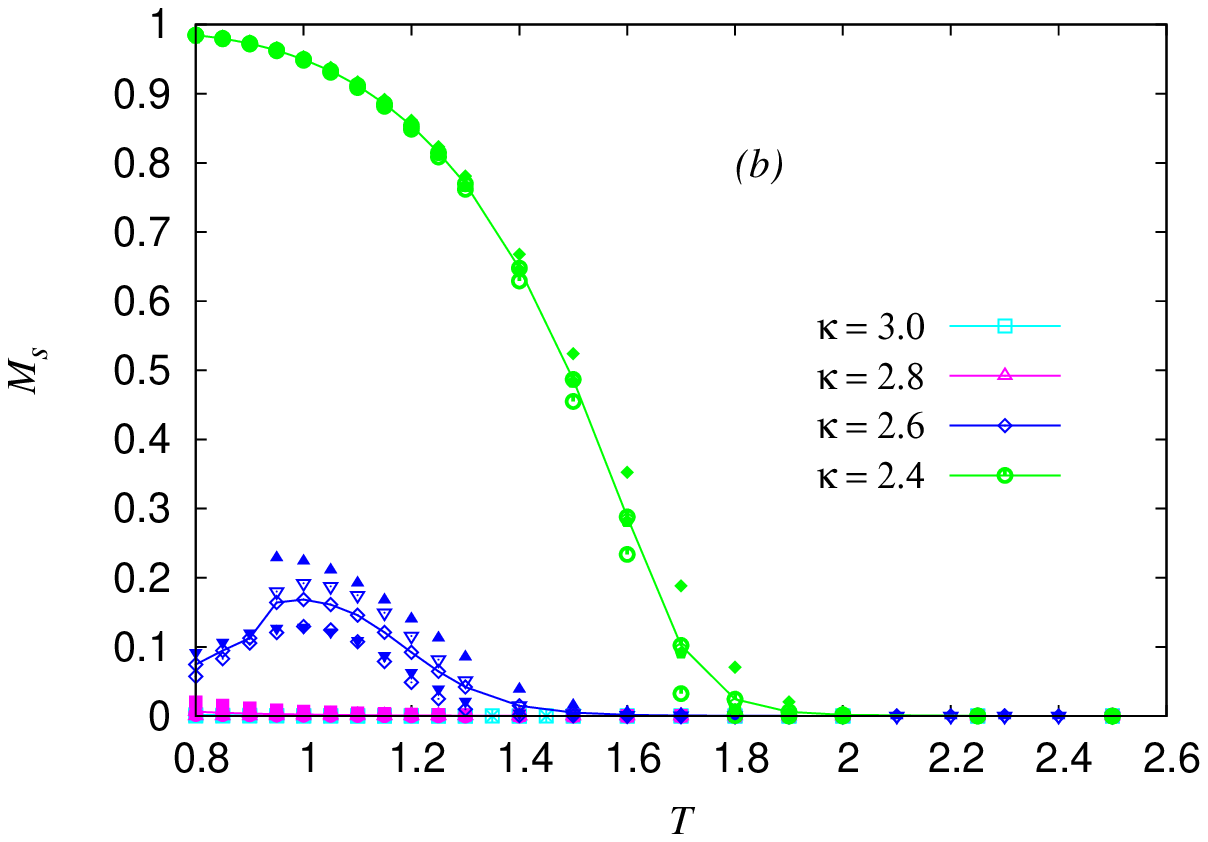}
\vspace{-0.1cm}
\caption{(a) Temperature dependence of Edward-Anderson parameter $q_{EA}$ and (b) Staggered 
magnetization $M_s$ of lattices with different values of $\kappa$. 
As shown, for $\kappa=2.4$, both quantities significantly increasing as temperature
decreasing.  The finite value of $M_s$ indicates a spatially ordered phase,  
signifies the absence  of SG phase (frozen random orientation).}
\label{qea}
\end{figure}

The temperature dependence of $q_{EA}$  
for various system sizes, $L = 16, 24, 32$ and  $48$ is shown in Fig. \ref{qea}(a). 
This parameter is to search for the existence of  SG phase transition in randomly 
frustrated system..
We also calculated  and the order parameter for AF system, the staggered 
magnetization $M_s$, shown  in Fig. \ref{qea}(b).
The characteristics of the  order parameter which increases with the decrease in temperature
is the indication of phase transition. This trend is indicated by 
  Fig. \ref{qea}, where for 
lattice with $\kappa=2.40$, both  parameters, $q_{EA}$ and $M_s$ are increasing.
However, it is to be noticed that the increasing $q_{EA}$ 
can not be considered as an indication of presence of SG phase
in the case $M_s$ also increases.
SG phase which is frozen random spin orientation
can not co-exist with  N\'eel phase. 
Therefore, the incerasing values of both parameters indicates the existence
of the same ordered phase, namely the N\'eel phase.
The plot of $q_{EA}$ for larger value of $\kappa$, i.e, $\kappa > 2.40$, indicates that
this parameter increases with the decrease in temperature.
However, it tends to decrease as system sizes increase.
This exhibits that there is no SG phase at thermodynamic limit.
The role played by the randomness, induced by random connectivity,
although results in frustrated state, it fail to bring low temperature SG phase.

The absence of SG phase and the presence of AF phase at low temperature
can be explained from the perspective of Harris criterion\cite{Harris},
which describes the role played by randomness on second order phase transition.
This is exactly what observed here. The pure AF Ising model on square
lattice does experience second order phase transition, with diverging
specific heat. As the number of links increase, more precisely,
the degree of randomness, the existing phase transition disappear
at certain value of $\kappa$, i.e., around 2.7.
 The precise value of $\kappa$ which destroy the 
AF order phase at low temperature will be the subject of our next study and will be 
published elsewhere.  With this paper, we report that non-canonical SG phase 
transition is not found for Ising model  in the rewired square
lattices with $ 2\leq \kappa \leq 3$. 

\section{Summary and  Conclusion}
This research studied the AF Ising model on rewired square lattices with
various average number of links (ANOL), associated with
coordination number for regular lattices. 
The lattices are obtained by randomly adding one extra link
to each site of the original square lattice, lead the rewired
latticess to have fractional ANOL, symbolized as $\kappa$,
varied from 2 to 3.  The added link
is constrained to connect a site to one of its diagonal neighbors.
Because of rewiring, there exist
abundance of triangular units in the lattice, in which spins are frustrated due 
to AF couplings. Accordingly, the system under investigated
 is in the class of random system with geometrical
frustration. We probed the role played by the randomness
and frustration in affecting the existing second order
phase transition of the native square lattice with AF interaction.
We used Replica Exchange Monte Carlo method which is considered
to be very powerful in dealing with randomly frustrated systems such as SGs.
From the time series of energy, we extracted energy and calculated the specific
heat. We also calculated the SG order parameter (Edward-Anderson parameter) 
as well as the staggered magnetization, which is the order parameter 
for AF ordering.

We observed  a clear indication of the effect
of randomness on the existing AF phase is clear and
found no SG phase transition.
The remnant of AF phase remains upto certain value
of ANOL, i.e.,   $\kappa \sim 2.7$.
The  pure AF square lattice is associated with
$\kappa = 2.0$ (no extra links added) 
 while for $\kappa = 3.0$, each site has obtained one extra link.  
The current topic has  opened another direction of research,
i.e., to study the interplay between randomness rendered by
rewiring procedure and the partially frustrated state.
In addition, to find the lower dimension for the existence
of Ising SG phase on rewired lattices is still desirable.
It will be our next topic to probe, for example searching for
Ising SG phase on rewired cubic lattices.
\section*{Acknowledgment}
The author wishes to thank A. G. Williams, D. Tahir and M. Troyer for 
stimulating discussions.  The computation of this work was carried out using parallel
computing facility in the  Physics Department, Hasanuddin University and 
the HPC facility of The Indonesian Institute of Science. The work is financially supported by
PUPT Research Grant, FY 2017 from The Indonesian Ministry of Research, Technology and 
Higher Education. 

\end{document}